\documentclass[prl,twocolumn,showpacs,superscriptaddress,preprintnumbers,amssymb]{revtex4}
\usepackage{graphicx}
\usepackage{dcolumn}
\usepackage{bm}
\usepackage{wasysym}

\newcommand{\beq}{\begin{equation}}
\newcommand{\eeq}{\end{equation}}
\newcommand{\beqn}{\begin{eqnarray}}
\newcommand{\eeqn}{\end{eqnarray}}

\begin{document}
\title{Novel Algebraic Boson Liquid phase with soft Graviton excitations}
\author{Cenke Xu}
\affiliation{Department of Physics, University of California,
Berkeley, CA 94720}
\date{\today}

\begin{abstract}
A bosonic model on a 3 dimensional fcc lattice with emergent low
energy excitations, with the same polarization and gauge
symmetries as gravitons is constructed. The novel phase obtained
is a stable gapless boson liquid phase, with algebraic boson
density correlations. The stability of this phase is protected
against the instanton effect and superfluidity by self-duality and
large gauge symmetries. The gapless collective excitation of this
phase closely resembles gravitons, although they have a soft
$\omega\sim k^2$ dispersion relation. The dynamics of this novel
phase is described by new set of Maxwell equations. This phase
also possesses an intricate topological order, requiring 18
winding numbers to specify each topological sector.
\end{abstract}

 \maketitle

The algebraic spin liquid is of crucial importance in
understanding the various types of behavior of strongly correlated
electron systems. This is not only because it belongs to the
family of the spin liquid, which has played a central role in the
attempt to understand superconductivity in the cuprates
\cite{anderson1987}\cite{zou1987}, but also because of its special
gapless excitation, which usually resembles one of the most
fundamental particles in the universe, the photon
\cite{sondhi2003}\cite{hermele2004}\cite{wen2003}. Unlike other
types of spin liquids (for instance, the $Z_2$ spin liquid of the
Quantum Dimer model on a triangular lattice \cite{sondhi2000}, and
a spin-1/2 model on Kagom\'{e} lattice \cite{balents2002}), the
algebraic phase behaves like a critical point, as the correlation
functions of physical quantities (for instance, the spin density
correlation function) fall off algebraically. Therefore when a
relevant perturbation is turned on, it is supposed to drive the
system into an ordered phase. It has been shown that the nature of
many kinds of normal order can be understood in their mother
phase, the 2+1d algebraic spin liquid \cite{hermele2005}.

Since the algebraic spin liquid phase is critical, it is usually
vulnerable to all kinds of relevant perturbations. In 2+1d,
monopole proliferation gaps the system and causes a certain
crystalline order to form which breaks translation symmetry.
Models with QED gauge symmetry can only give rise to gapped gauge
bosons\cite{wen2003}. Monopoles in 2+1d can only be irrelevant due
to the Berry phase at the transition point between different types
of crystal order in the quantum dimer model \cite{rokhsar1988,
fradkin1990, henley1997}, or between crystal order and superfluid
order \cite{senthil2004, ashvin2004}. Protecting the algebraic
spin liquid phase is equivalent to protecting gapless bosonic
excitations without symmetry breaking. As is well-known,
continuous symmetry breaking is one way to obtain a gapless
bosonic excitation. The gapless excitation is the Goldstone mode
\cite{goldstone}. However, the algebraic spin liquid which is a
phase without any symmetry breaking, requires a different
mechanism to guarantee the stability of its gaplessness and
criticality. So far, the only stable algebraic spin liquid phase
that has been proven to exist is the 3+1d photon spin liquid. This
spin liquid can be realized in either the quantum dimer model on
the cubic lattice \cite{sondhi2003}, or a spin-1/2 model on the
pyrochlore lattice \cite{hermele2004}. In a 2+1d system, monopole
proliferation is a concern, although the existence of a 2+1d
stable algebraic liquid phase has been argued, by coupling the
compact gauge fields with a large number of flavors of the matter
field \cite{hermele2004a}.

In this paper, a new type of algebraic liquid phase is shown to exist. Since
a spin system can usually be mapped to a bosonic
system through the identification, $n_i - \bar{n} = s^z_i$ and
$b^\dagger_i \sim s^+_i$, the bosonic language is used in this paper. A
lattice model is constructed, which shows a stable algebraic boson
liquid phase with gapless collective excitations. The gapless
collective excitations have the same polarization and gauge
symmetry as the graviton, although the dispersion relation is softened
to $\omega\sim k^2$. The stability of this algebraic phase is due
to the self-duality and large gauge symmetry of the low energy
effective Hamiltonian. This work focuses on the basic
properties of this new phase. We leave the discussion of the 2d
case and the transition to other phases to another piece of work
\cite{xu2006}.

Let us first briefly review some basic properties of the 3d photon
liquid phase as a warmup. Although many different models have been
proposed in the last few years
\cite{sondhi2003}\cite{hermele2004}\cite{wen2003}, the phases
obtained have very similar properties. Therefore, let us take a 3d
Quantum Dimer Model on the cubic lattice as an example. On the cubic
lattice, every site is shared by six links, and only one of those
links is occupied, by exactly one dimer. On every
square face, if two parallel links are occupied by
dimers, they can be resonated to dimers perpendicular
to the original ones. Besides this resonating term, another
diagonal weight term for each flippable plaquette is also included
in the Hamiltonian \beqn H = \sum -t(| \parallel \rangle\langle =
| + h.c.) + V(| \| \rangle\langle \| | + | = \rangle\langle =
|)\label{dimer}\eeqn

The dimer model can be mapped onto a rotor model. A rotor number can
be defined on each link to describe the presence or absence of
dimers: $n = 1$ when the link is occupied by a dimer, and $n = 0$
when the link is empty. The Hilbert space of this quantum system
is a constrained one, with the constraint $\sum_{i = 1}^6 n_i = 1$
around each site. Let us define the quantity $E_{i,\hat{a}}$ as
$E_{i,\hat{a}} = (-1)^{i_x+i_y}n_{i,\hat{a}}$. Now, the constraint
of this system can be rewritten as the Gaussian law for electric
fields, $\partial_\mu E_\mu = \pm 1$. Notice that because the
quantity $E_{i,\hat{a}}$ is defined on links, a natural vector
notation can be applied: $\vec{E}_i = (E_{i,\hat{x}},
E_{i,\hat{y}},E_{i,\hat{z}})$.

The background charge distribution $\pm 1$ on the cubic lattice
plays a very important role in the solid phase, i.e. the confined
phase. The form of the crystalline phase can be determined from
the Berry phase induced by the background charge distribution
\cite{ashvin2004}\cite{hermele2004}. However, as we are focusing
on the algebraic liquid phase, the background charge is not very
important. Let us instead impose the constraint $\partial_\mu
E_\mu = 0$. Because of this local constraint, the low energy
physics will be invariant under the gauge transformation $\vec{A}
\rightarrow \vec{A} + \vec{\nabla}f$ ($\vec{A}$ is the conjugate
variable of $\vec{E}$), which is exactly the gauge symmetry of the
$U(1)$ gauge theory. Now the effective Hamiltonian of this theory
should be invariant under this gauge transformation. The
Hamiltonian (\ref{dimer}) can be effectively written as \beqn H =
\sum - \tilde{t}\cos(\nabla\times\vec{A}) + 1/(2\kappa)\vec{E}^2
\eeqn This Hamiltonian is the 3+1d compact QED, which should have
a deconfined photon phase (\cite{polyakov1977,polyakov1987}). The
reason for the existence of the photon phase is actually the
self-duality and gauge symmetry, as discussed below.

First, because the constraint $\partial_\mu E_\mu = 0$ is strictly
imposed on this system, the matter field, i.e. the defect which
violates this constraint, is absent. As is well known, the $U(1)$
gauge symmetry cannot be broken spontaneously without a matter
field \cite{fradkin1978}. Therefore the dimer superfluid order is
ruled out. The other instability of the photon phase is towards
the gapped solid phase, which can be analyzed in the dual theory.
We can introduce the dual vector $\vec{h}$ and dual momentum
vector $\vec{\pi}$ ($h$ and $\pi$ are both defined on the faces of
the cubic lattice) as $\vec{E} = \nabla\times \vec{h}$ and
$\nabla\times \vec{A} = \vec{\pi}$. One can check the commutation
relation and see that $\vec{h}$ and $\vec{\pi}$ are a pair of
conjugate variables. The Gaussian law constraint on this system is
automatically solved by the vector $\vec{h}$. Now, the field
theory for the photon phase is self-dual: \beqn H = -
t_1(\nabla\times\vec{A})^2 + c_1\vec{E}^2, \nabla\cdot\vec{E} =
0,\cr = - t_2(\nabla\times\vec{h})^2 + c_2\vec{\pi}^2,
\nabla\cdot\vec{\pi} = 0. \eeqn In principle, a vertex operator
$\cos(2\pi N\vec{h})$ is supposed to exist in the dual
Hamiltonian, due to the fact that $\vec{E}$ only takes on integer
values. $N$ is an integer corresponding to the Berry phase of the
vertex operator. When this vertex operator is relevant, it will
gap out the photon excitation and drive the system into a
crystalline phase, according to the Berry phase. However, the
vertex operator is irrelevant in the photon phase. Notice that,
because the theory is self-dual, the dual theory has the same
gauge invariance $\vec{h} \rightarrow \vec{h} + \vec{\nabla} f$ as
the original theory. However, the vertex operator $\cos(2\pi
N\vec{h})$ breaks gauge invariance and thus the correlation
function between two vertex operators in this photon phase is
zero, i.e. the vertex operator is irrelevant in this Gaussian
theory. Unless the dual charges (the monopole) condense and break
the dual $U(1)$ gauge symmetry, the photon phase is always stable.
The gaplessness of this phase is protected by both self-duality
and the gauge symmetry.

The photon phase is an algebraic liquid phase, the dimer
density-density correlation function falls off algebraically. When
$t = V$ in (\ref{dimer}), the lattice model can be solved exactly,
and the equal-time density correlation falls off as $\langle
n(0)n(r) \rangle \sim 1/r^3$ \cite{sondhi2003}.

\begin{figure}
\includegraphics[width=2.7in]{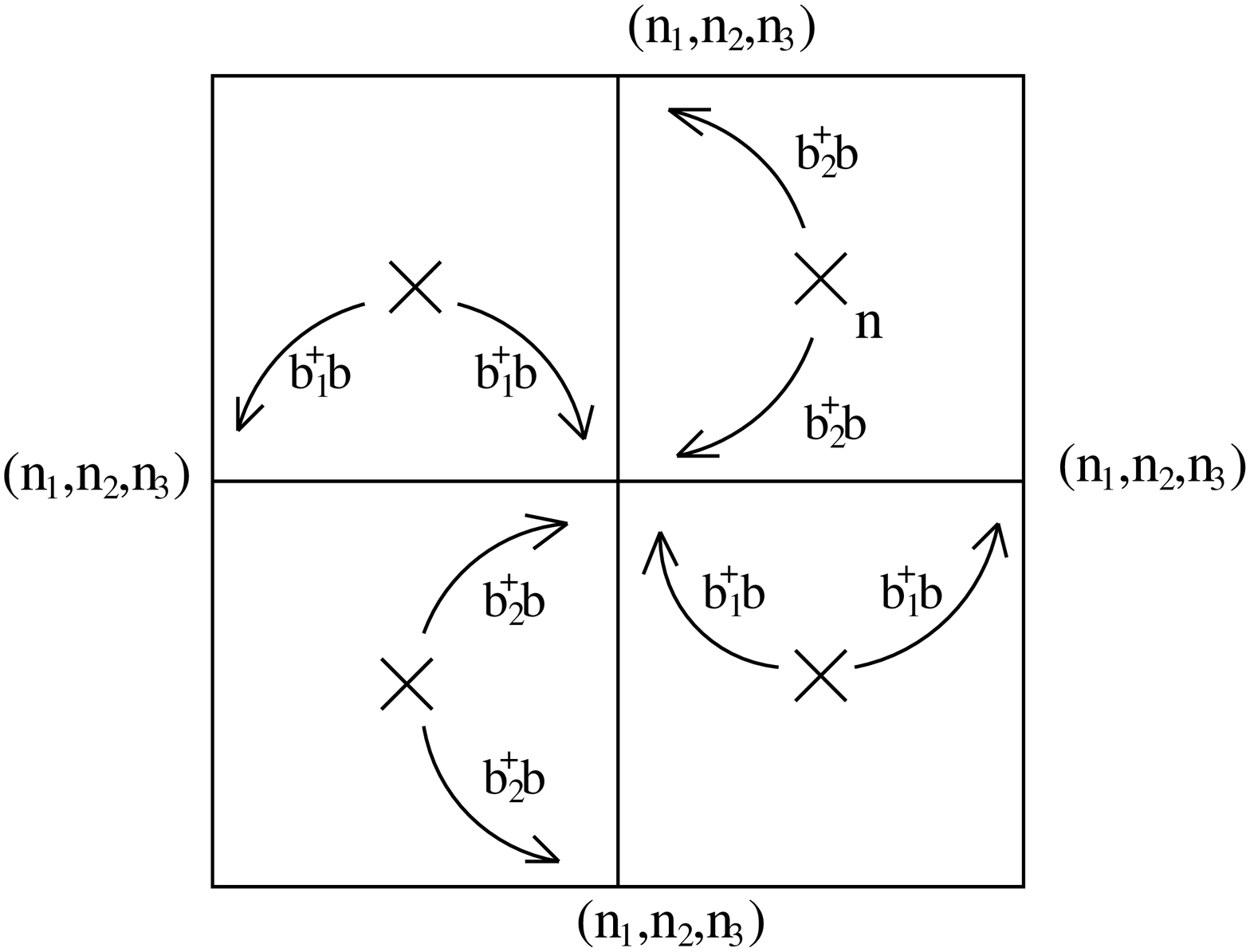}
\caption{The $xy$ plane of the fcc lattice. On each site there are
three orbital levels, we denote the particle numbers as $(n_1,
n_2, n_3)$; on each face center there is one orbital level, we
denote the particle number as $n$.} \label{lattice}
\end{figure}

Let us now move on to our graviton model. A 3d fcc lattice is
considered and physical quantities are defined on both the sites
and face centers of the cubic lattice. Let us assume there are 3
orbital levels on each site, and 1 orbital level on each face
center. The particle number on the face center is denoted as $n$,
and the particle numbers on sites are denoted as $(n_1,n_2,n_3)$ in Fig.\ref{lattice}.

The Hamiltonian for this system contains three parts, $H = H_0 +
H_1 + H_2$. $H_1$ is the nearest neighbor hopping between sites
and their nearest face centers, and also between adjacent face centers (notice
that adjacent face centers have the same distance between them
as a site and its nearest face center). $H_1 = (\sum_{<i,\bar{j}>}\sum_{a
= 1}^3-tb^\dagger_{a,i}b_{\bar{j}} - \sum_{<\bar{i},\bar{j}>}t
b^\dagger_{\bar{i}}b_{\bar{j}}+h.c.)$, $H_2$ is the on site
interaction $H_2 = U(n - 1)^2$, which fixes the average filling of
the fcc lattice. $H_0$ is the interaction term involving links in
all 3 directions. For example, for the link in $(i,\hat{x})$, the
interaction term reads \beqn H_0 = V(n_{i + 1/2\hat{x} +
1/2\hat{z}} + n_{i + 1/2\hat{x} - 1/2\hat{z}} + n_{i + 1/2\hat{x}
+ 1/2\hat{y}} \cr + n_{i + 1/2\hat{x} - 1/2\hat{y}} + 2n_{1,i} +
2n_{1,i+\hat{x}} - 8)^2 \label{h0}\eeqn. The links in $\hat{y}$ and $\hat{z}$ directions are treated similarly. If the bracket in
(\ref{h0}) is expanded, it becomes the usual two body repulsion
term.

\begin{figure}
\includegraphics[width=2.0in]{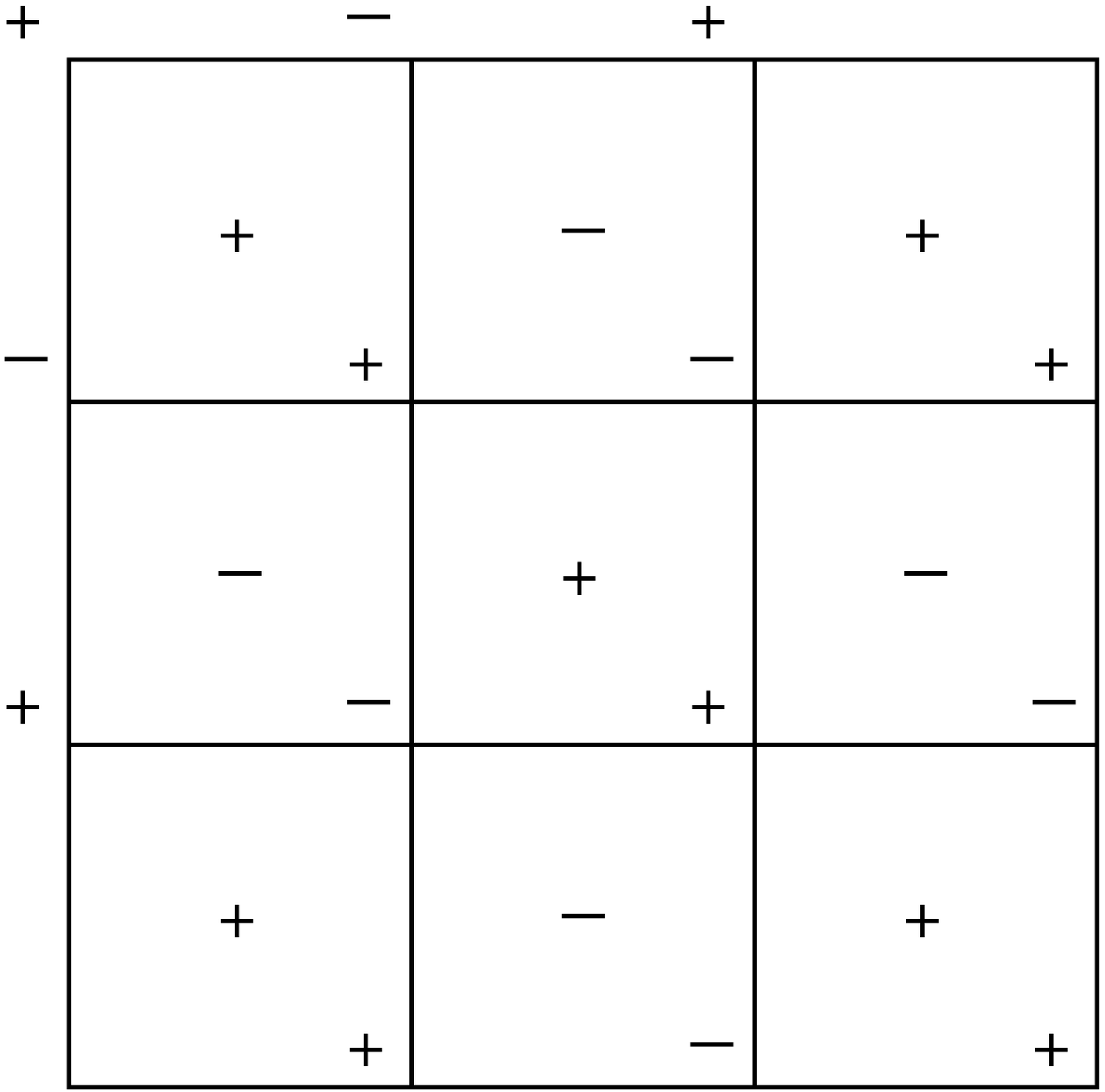}
\caption{The distribution of the sign $\eta$ on $xy$ plane. After
introducing $\eta$, the constraint can be written as in
(~\ref{cons3}).} \label{signgrav}
\end{figure}

When $H_0$ becomes the dominant term in the Hamiltonian, it
effectively imposes a constraint on the system. The best way to
view this constraint is by introducing a staggered sign and defining
new variables, similar to the electric field in the dimer model
discussed before. Let us define a rank-2 tensor $E_{ij}$. The
off-diagonal terms are defined on face centers as
$E_{ij} = \eta_r (n - 1)$. $n$ is located at one of
$\hat{i}\hat{j}$ face centers; the diagonal term is defined on
sites as $E_{ii} = \eta_r (n_{i} - 1)$ with $i = 1,2,3$. $\eta_r =
\pm 1$ and the distribution of sign $\eta_r$ is shown in Fig.
\ref{signgrav}

After introducing the sign $\eta$, the constraint effectively
imposed by (\ref{h0}) can be compactly written as \beqn
2\partial_{x}E_{xx} + \partial_y E_{xy} +
\partial_{z}E_{xz} = 0, \cr \partial_{x}E_{xy} + 2\partial_y E_{yy} +
\partial_{z}E_{yz} = 0, \cr \partial_{x}E_{xz} + \partial_y E_{yz} +
2\partial_{z}E_{zz} = 0.\label{cons3}\eeqn A violation of this
constraint can be interpreted as a charged defect excitation and
can drive the system into an ordered boson superfluid state after
condensation. We will discuss this transition in another paper
\cite{xu2006}. In the current work we only focus on the case when
the constraint is strictly imposed. The constraint (\ref{cons3})
requires the low energy Hamiltonian to be invariant under the
gauge transformation $A_{ij} \rightarrow A_{ij} + \partial_if_j +
\partial_jf_i$ . This is precisely the gauge symmetry of the
graviton in 3d space. Thus, the low energy physics can only
involve the gauge invariant quantity $ R_{\alpha\mu\beta\nu} =
1/2(A_{\alpha\nu,\mu\beta} + A_{\mu\beta,\alpha\nu} -
A_{\mu\nu,\alpha\beta} - A_{\alpha\beta,\mu\nu})$, the linearized
curvature tensor. Here the convention of general relativity has
been used: $A_{\alpha\nu,\mu\beta} =
\partial_{\mu}\partial_{\beta} A_{\alpha\nu}$.

In 3d space, the curvature tensor has 6 nonzero components,
because the total number of nonzero components is reduced by the
symmetry of curvature tensor $R_{ijkl} = - R_{jikl} = - R_{ijlk} =
R_{klij}$ \cite{wheeler1973}. Thus, now the effective low energy
Hamiltonian reads \beqn H_{ring} = \sum_{ij,i\neq
j}-\tilde{t}_1\cos(R_{ijij}) - \sum_{ijk,i\neq j, j\neq k, i\neq
k} \tilde{t}_2\cos(R_{ijik}) \cr + \frac{1}{2\kappa_1}(\sum_{i =
1}^3E^2_{ii}) + \frac{1}{2\kappa_2}(\sum_{ij,i\neq
j}E_{ij}^2)\label{ring}\eeqn The cosine terms involving the
curvature tensor are ring exchange terms generated by the nearest
neighbor hopping, resembling the dimer flipping term
$\cos(\nabla\times\vec{A})$ in (\ref{dimer}). Ring exchanges in
(\ref{ring}) are generated at the eighth order perturbation of the
nearest neighbor hopping, $\tilde{t}_1$, $\tilde{t}_2\sim
t^8/V^7$. All lower order perturbations only generate terms which
do not comply with the constraint (\ref{cons3}). One of the ring
exchanges is shown in Fig. \ref{lattice} and it corresponds to the
term $\cos(R_{xyxy})$.

In order to derive the correct action, one needs to introduce
a Lagrange multiplier $A_{i0}$ for the constraint (\ref{cons3}). The
full action reads \beqn L = \sum_{i,j}(\partial_{\tau}A_{ij} -
\partial_iA_{j0} - \partial_{\tau}A_{i0})^2 +
\sum_{ij,i\neq j}\tilde{t}_1\cos(R_{ijij}) \cr + \sum_{ijk,i\neq
j, j\neq k, i\neq k} \tilde{t}_2\cos(R_{ijik}) \eeqn Now, the
gauge symmetry can be enlarged to quantities defined in
space-time: $A_{\mu\nu} \rightarrow A_{\mu\nu} +
\partial_\mu f_\nu + \partial_\nu f_\mu $ and $A_{00} = 0$, $f_{0} =
0$. In this system, the boson superfluid order is again ruled out
by the gauge symmetry. Without crystalline order (proven later),
the system is in a liquid phase with excitations which have the
same gauge symmetry (and hence the same polarization) as the graviton.
Unlike the quantum dimer model, the curvature tensor is a second
spatial derivative of $A_{ij}$ and the graviton mode has a soft
dispersion $\omega \sim k^2$.

Whether the graviton excitations survive, or crystal order
develops can be determined in the dual theory. If we define
the symmetric tensor $\mathcal{E}_{ij}$ as  $\mathcal{E}_{ii} =
\sqrt{2}E_{ii}$, $\mathcal{E}_{ij} = 1/\sqrt{2}E_{ij}, i \neq j$,
the constraint (\ref{cons3}) can be solved by defining the dual tensor
$h_{ij}$ as $ \mathcal{E}_{ij} =
\epsilon_{iab}\epsilon_{jcd}\partial_a\partial_ch_{bd}$. This is a
double curl of the symmetric tensor $h_{ij}$. $h_{ij}$ also lives on
the sites and faces of this fcc lattice.

If checked carefully, one can notice that, the curvature tensor
can also be written in the double curl form  \beqn 2R_{xyxy} =
\epsilon_{zab}\epsilon_{zcd}\partial_a\partial_cA_{bd}, 2R_{xzxz}
= \epsilon_{yab}\epsilon_{ycd}\partial_a\partial_cA_{bd} ,\cr
2R_{yzyz} =
\epsilon_{xab}\epsilon_{xcd}\partial_a\partial_cA_{bd}, 2R_{xyxz}
= \epsilon_{yab}\epsilon_{zcd}\partial_a\partial_cA_{bd},\cr
2R_{yxyz} =
\epsilon_{xab}\epsilon_{zcd}\partial_a\partial_cA_{bd}, 2R_{zxzy}
= \epsilon_{xab}\epsilon_{ycd}\partial_a\partial_cA_{bd}.
\label{doublecurl}\eeqn Therefore, this model is precisely
self-dual, as long as we define the dual variables $h_{ij}$ in
terms of $ \mathcal{E}_{ij} =
\epsilon_{iab}\epsilon_{jcd}\partial_a\partial_ch_{bd}$ and its
conjugate $\pi_{ij}$ as follows: \beqn \sqrt{2}R_{xyxy} =
\pi_{zz}, \sqrt{2}R_{yzyz} = \pi_{xx}, \sqrt{2}R_{xzxz} =
\pi_{yy}, \cr 2\sqrt{2}R_{xzyz} = \pi_{xy}, 2\sqrt{2}R_{xyxz} =
\pi_{yz}, 2\sqrt{2}R_{xyzy} = \pi_{xz}.\eeqn $\pi_{ij}$ is subject
to the same constraint as $E_{ij}$.

Now the dual action reads \beqn L_{dual} =
\sum_{ij}(\partial_th_{ij} - \partial_ih_{j0} -
\partial_jh_{i0})^2 - \sum_{ij,i\neq j}\rho_1\tilde{R}^2_{ijij}
\cr - \sum_{ijk,i\neq j,j\neq k, i\neq k}\rho_2\tilde{R}^2_{ijik}
+ \cdots\cdots \label{dualac3}\eeqn $\tilde{R}_{ijkl}$ is the
curvature tensor of $h_{ij}$. $h_{i0}$ is a Lagrange multiplier, due
to the constraint on $\pi_{ij}$. The ellipses include possible
vertex operators. Without the vertex operators, this theory is at
a Gaussian fixed point and hence in an algebraic liquid phase with
soft graviton excitations. If the vertex operators are relevant,
they will destabilize the liquid phase and gap the graviton
excitation, and form crystalline order according to the Berry phase.
The dual action (\ref{dualac3}) is also invariant under
the gauge transformation $h_{\mu\nu} \rightarrow h_{\mu\nu} +
\partial_\mu f_\nu +
\partial_\nu f_\mu$, with $h_{00} = 0$ and $f_0 = 0$.
Therefore, any kind of vertex operator (for example $\cos(2N\pi
h_{ij})$) is not a gauge invariant operator. The correlation
function between two vertex operators at the Gaussian fixed point
is zero or correlated at very short range, so the Gaussian fixed
point (also the algebraic spin liquid) is stable against weak
perturbations of the vertex operators.

In the critical liquid phase, correlation functions between local
order parameters decay algebraically. In our case, the
correlation function decays more rapidly than in the QED model, because
the local order parameter is the second order derivative of the dual
variable $h_{ij}$. For instance, the correlation between two
spatially distant boson density operator $n_1 - \bar{n}$ scales as
$\sim (-1)^{x+y+z}/ r^5$.

The dynamics of the liquid phase can be described by a new set of
Maxwell equations. Define the rank-2 tensor $\mathcal{B}_{ij} =
\epsilon_{iab}\epsilon_{jcd}\partial_a\partial_c A_{bd}$. The
dynamical equations that describe this liquid phase are \beqn
\partial_i\mathcal{E}_{ij} = \partial_i\mathcal{B}_{ij} = 0,
\cr \partial_t\mathcal{E}_{ij} -
\kappa\epsilon_{iab}\epsilon_{jcd}
\partial_a\partial_c\mathcal{B}_{bd} = 0,\cr
\partial_t\mathcal{B}_{ij} + \kappa\epsilon_{iab}\epsilon_{jcd}
\partial_a\partial_c\mathcal{E}_{bd} = 0.\label{maxell}\eeqn Charged excitations
and topological defects are absent in these equations, so they
correspond to the Maxwell's equations in vacuum. If the constraint
(\ref{cons3}) is softened, charge density and charge current have
to be incorporated in equation (\ref{maxell}). Because the field
variables are rank-2 tensors, the charge density will be a vector
field, and the charge current will again be a rank-2 tensor.
Therefore, the form of charge density and current are very similar
to spin density and spin current. We will leave the detailed
discussion to the other paper \cite{xu2006}.

Since the spin liquid does not break any symmetry, one cannot
classify spin liquids by symmetry. Instead, topological order has
been introduced to classify different spin liquids \cite{wen1991}.
The topological sector of the spin liquid phase with photon
excitations can be described by 6 integers, 3 of them describe
electric flux through the system and the other 3 describe the
magnetic flux \cite{hermele2004}. In our case there are more
topological sectors. Consider a $yz$ plane in the original lattice
and sum over $E_{xx}$ of a particular configuration on this plane.
This gives us an integer. This integer is invariant under any ring
exchange allowed. In addition, this quantity does not depend on
which $yz$ plane is chosen, this is due to the constraint
(\ref{cons3}) and Gauss's theorem, $\int d^3x
\partial_ie_{ix} = \int e_{ix}\cdot d S_i = 0$. The
total integral of $E_{ix}$ on any $yz$ plane gives the same
result. Let us denote this integral as $n_{E,xx,yz}$.

There are similar integrals which also specify a topological
sector: they are $n_{E, yy, xz}$, $n_{E, zz, xy}$, $n_{E, xy,
yz}$, $n_{E, xy, xz}$, $n_{E, yz, xz}$, $n_{E, yz, xy}$, $n_{E,
xz, xy}$, $n_{E, xz, yz}$. There are in total 9 types of electric
flux winding numbers. The dual field $\pi_{ij}$ is subject to the
same type of constraint as (\ref{cons3}), because of the
self-duality. Thus, the flux of $\pi_{ij}$ provides another 9
integrals specifying a topological sector. In total there are 18
winding numbers.

To conclude, in this work a new type of stable algebraic boson
liquid phase has been realized in a 3d lattice model. Low
temperature physics should be controlled by the gapless graviton
excitations, for instance, the gravitons make the contribution to
the specific heat $C \sim T^{3/2}$, which is larger than the $T^3$
contribution from phonons. If the original model was written in
terms of fermionic particles, novel non-Fermi liquid behavior is
expected.

The author would like to thank L. Balents, M. P. A. Fisher, A.
Vishwanath, J. E. Moore and C. Wu for helpful discussions.

\bibliography{gravitonliquid}

\end{document}